# Media Objects in Time - A Multimedia Streaming System - Work in Progress Paper v 1.5 - [†]

*Björn Feustel, Thomas C Schmidt*
{feustel,schmidt}@fhtw-berlin.de

Computer Centre, Fachhochschule für Technik und Wirtschaft Berlin
Treskowallee 8, 10318 Berlin, Germany.

**Abstract**

*The widespread availability of networked multimedia potentials embedded in an infrastructure of superior quality gives rise to new approaches in the areas of teleteaching and Internet presentation: The distribution of professionally styled multimedia streams is within the realm of possibility. This paper presents a prototype - both model and runtime environment - of a time directed media system treating any kind of presentational contribution as reusable media object components. The plug-in free runtime system is based on a database and allows for a flexible support of static media types as well as for easy extensions by streaming media servers. The prototype implementation includes a preliminary version of Web Authoring platform.*

## 1 Introduction

Today's standards of internet-connected computers provide associatively linked texts, images, sounds, movies or other information material. They thereby confront students as well as teachers with a new paradigm of knowledge transfer. Networked multimedia accessories not only transport a formerly unknown multitude of presentation methods to the lecture hall or, in the framework of teleteaching, to students homes but they also present an unfiltered totality of present knowledge, with which, in terms of quantity and rapidity of change, no single teacher can compete.

However, individual aspects of classroom communication remain bound to traditional lecture forms. These are, first of all, direct dialogue and related interchanges. Secondly, and almost equally important is the notion of time and speed that teacher imposes on his students by determining the order and pace in which the material is rolled out and finally by fixing the dates for testing and for documenting success. It is this time control process that ultimately determines performance.

Many attempts are made to use World Wide Web techniques to at least partially substitute traditional lectures. The intention of this is to liberate students from the restrictions of temporal and spatial presence and thereby offer knowledge to a wider public. But in contrast to a personal classroom attendance, an http-based training module is driven by end user's mouse clicks and thus delegates the ordering and timing completely to the recipient of the learning material. Therefore it places on the learner the responsibility not only for overall understanding but also for undertaking each progressive step in time.

[†] This work was supported in part by the EFRE program of the European Commission.

The important concept of time in teaching is one major reason for all the attention in recent research works to World Wide Web techniques that distribute multimedia documents with temporal and spatial relations. The growing demand for synchronised handling of time-based media such as video and audio serves as a second motive for introducing temporal aspects to the Web. Finally, streaming data sources create a new level of scalability by accounting for transport timing and therefore become important quantitatively throughout the Internet.

In the present paper we present the project Media Objects in Time (MobIT) that develops a model for time-directed presentation and processing of universal media objects, as well as a prototype for a MobIT runtime environment. The open system operates on standard formats for multimedia information content. Structural relations between documents are specific to the model, but may be easily extracted from the underlying XML type definition.

This paper is organised as follows. In section 2 we introduce the basic ideas and functionality of our project and draw comparison to related work. Section 3 presents the underlying Compound Flow Model (CFM). Architecture and implementation properties of the MobIT runtime environment will be discussed in section 4. Finally, section 5 is dedicated to conclusions and an outlook on the ongoing work.

## 2 MobIT and related works

### 2.1 Presenting a media stream

The teaching and presentation system introduced here centres around the idea of media objects that can be synchronised in time and that can be linked to form fairly complex presentations. But at the same time, any object remains self-consistent and can be used independently. Roughly speaking, our basic concept consists of defining media object instances and lining them up in time as is shown in figure 1. MobIT intends to provide an accurate scheme for temporal and spatial placement of presentation objects, where authors do not have to take care of interobject synchronisation dependencies or adaptation to possibly inaccurate network performance, the latter being subject to implementation of latency hiding techniques.

Presenting itself on a timeline, each presentation becomes a time-based data object, even if composed only of timeless media such as texts or images. Any presentation component will carry an instance of initial appearance and a moment (possibly at infinitum) for fading away from the client's screen. Within this framework, any streaming media such as video or audio may be included

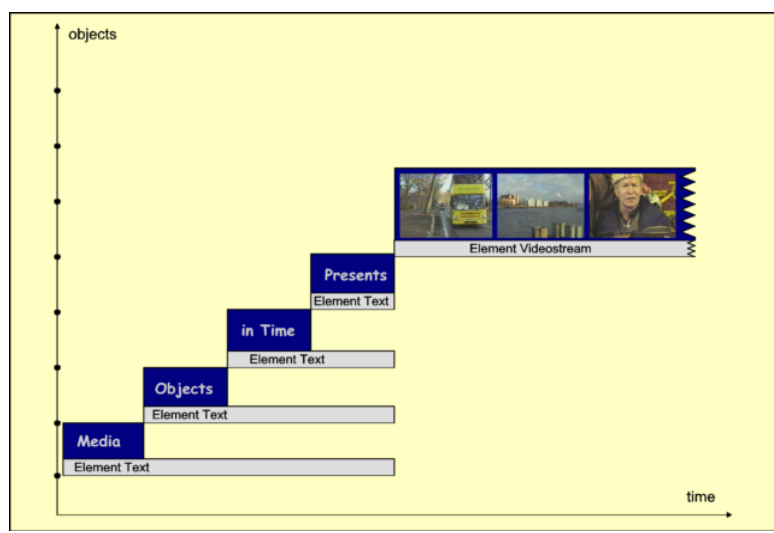

**Figure 1: Object Instances aligned to time**

and synchronised to the scene and the overall data stream.

Aiming at combined utilisation in lecture rooms as well as in teleteaching, our model focuses on a clear, straightforward concept of reusable compound media components. Screenplay scripts arranging their behaviour in space and time will accompany these. Thus, instead of the page-oriented WWW concept or the typically event-driven nature of CBT products, MobIT runs as a flow-oriented presentation model showing, for example, a crash-test video combined subsequently with charts of relevant statistics and vocally explained CAD car models.

The implementation of the MobIT runtime prototype concentrates on a pure JAVA solution. Any recipient may load the corresponding JAVA applet into the browser, which will then connect to the MObIT server, request for meta-data, open media data streams and eventually start the timer and the display process. Spatial and temporal placement is done with great accuracy. Additional plug-in software is not required.

## 2.2 Related work

As mentioned earlier, several interesting research activities presently impose the concept of time to the Web, the most prominent being the W3C recommendation Synchronized Multimedia Integration Language (SMIL) [4]. As a declarative language, SMIL allows for synchronisation of media objects in a somewhat simplistic, HTML-style fashion. Synchronisation is done in object pairs, either sequentially or in parallel. The appearance of any object may be bound to a duration parameter. SMIL extends the notion of hyperlink to connecting temporal and spatial sub regions.

The runtime behaviour of a SMIL interpreter is thereby left more or less open, which is probably the most important drawback of the model. Combined with the absence of a stringent handling of timelines, temporal inconsistencies in more complex documents can be foreseen. Besides few reference implementations of SMIL players, there is an attempt to include synchronisation features into the Web browser called HTML+TIME [5]. This proposal addresses temporal extensions to HTML and incorporates basic elements of SMIL.

However both ideas, suffer substantial limitations due to the simplistic design of HTML, which has no structuring for media object use. Rutledge et al [7] consequently report about severe difficulties in authoring SMIL presentations mainly due to the lack of reusability of object compositions, as well as to SMIL's inability to deal with complex object relations. In the most recent work, the Boston specification of SMIL [6], the World Wide Web Consortium heads for a realisation of SMIL as a module within the framework of the XHTML language. Most of this work is presently ongoing and far too incomplete to allow implementation.

As a completely different example more similar to our work, we would like to mention the Nested Context Model (NCM) of Soares et. al. [8]. With the aim of establishing a strong structure for flexible deployment of hypermedia documents, the NCM provides composite meta structuring for different media types, called nodes, up to an arbitrary level of complexity. These nodes may contain a reference list of denoted nodes, giving rise to an arbitrary graph structure of the composed document. The model, which has been implemented in a system called HyperProp, treats hypermedia documents essentially as passive data structures. Synchronisations are defined through events, which may occur as the result of object presentation or user interaction.

Since embedding of media objects within the NCM results in a passive mesh without further presentational meaning, an additional structure of activation, events and contexts (called perspectives), has to be superimposed. This characteristic on the one hand leaves some freedom to the author (the same object structure may encounter different behaviour in different contexts); on the other hand it adds an additional level of complexity to the modelled hypermedia system, which represents the major difference from our work.

# 3 The Compound Flow Model

## 3.1 Constituents

In designing the Compound Flow Model (CFM) much care was taken to define a simple structure of straightforward logic, which intuitively appeals to document authors and is suited for lightweight implementation, especially on the client side. The CFM centres on a uniform hull entity called Media Object (Mob) serving as a universal container for embedding either subsequent Mobs or media data. By referencing one another, Mobs allow for hierarchical compositions of unlimited depth, where the leaves of the resulting tree structure are fed with the actual media information by means of a distinct data object called Element (s. figure 2).

Whereas Mobs as uniform composition objects exhibit neutral behaviour with respect to embedded data, Elements are of specific, atomic type and enclose all properties related to the media data they control. Information about the element type is exchanged via the MIME-type standards according to RFC 1341 and is used by the element class to appropriately receive media data, compose presentation methods and finally display and delete them on the screen. This data processing of elements applies for static media such as images or text, as well as for sub served data like video or audio streams.

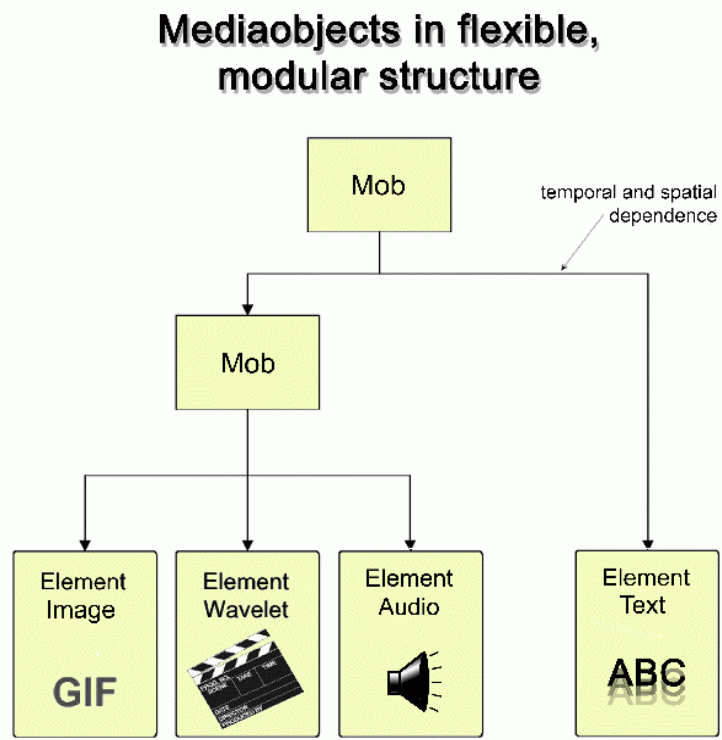

Figure 2: Media Object Hierarchy

Vital to the framework of CFM is an environment for generating and controlling the flow. As media objects for a given presentation may be widely branched, each one of them equipped with a complex structural inheritance and its own synchronisation demands, a flow control module needs to resolve all structural data dependencies. It thereafter has to linearise resulting bulk information, to form an ordered flow and finally add objects to the externally provided primary timer.

## 3.2 Media Objects

Media objects may be seen as the central constituents making up the CFM data structure. As the basic design idea, a Mobs consists of both the subordinate object reference list and a screenplay script for the references, describing all parameters responsible for their behaviour in time and space. These scripts we denote as Playlists. Playlists describe the total states attained by the corresponding Mobs.

The notion of generalized reusability of any of the components involved is tightly bound to the concept of combined reference to objects and their states. Roughly speaking, an object exhibits generalised reusability if and only if it is self-consistent and parameterisable in state space. The fundamental parameters of the state space up to now are the spatial size and the duration in time. Some additional features such as background colour or change of font type have been implemented. However, the definition of sustainable parameters is still somewhat vague. The actual realization of parameter values needs to be done at the level of individual Elements, e.g. scaling, clipping, ... Nevertheless any Mob is equipped with the ability to receive and process available parameters.

Self-consistency of Mob hierarchies may be viewed either from a structural or a state space perspective; structural self-consistency is present as long as the object reference hierarchy omits recursions, i.e. a Mob must not contain any referral pointing to itself or a subordinate object. Self-consistency in state space is ensured as long as any Mob stays within the temporal and spatial bounds of its superMob.

To strengthen this two-fold notion, our Mob design includes the congruence of consistency in structure and state space: As an integral part of the CF model, structural arrangements of Mobs carry the notion of inclusion relations in space and time. As one result, scale parameters are processed along the object hierarchy to hand down actual Element sizes. The exact appearance of a Media Object thus not only depends on its individual referenced parameters but also on its relative position within the hierarchy. As one benefit, it should be noted that an author can place and parameterise any compound object in any context and the system will ensure spatial and temporal integrity.

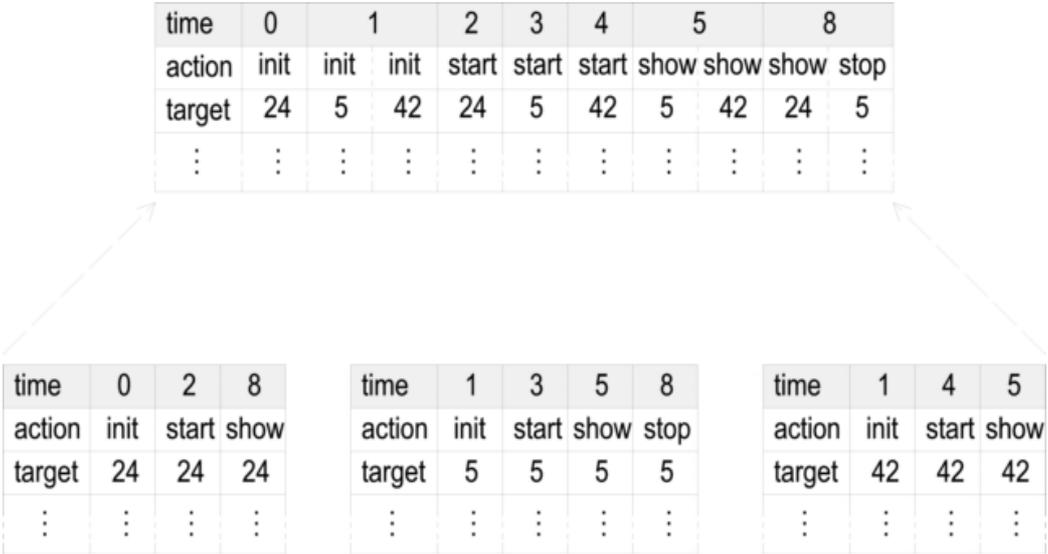

Figure 3: : Playlists under Linearisation

### 3.3 The compound flow

As is clear from the above, every Mob provides its own relative co-ordinate system in space and time, each of them needing transformation into display co-ordinates during runtime. Reference identifiers of objects rely on a scheme of global database numbering and need to be transferred into local presentation co-ordinates as used in [3]. Most importantly, Mobs are structured in a possibly complex, highly non-linear fashion that may be suitable for spatial representation but is inappropriate for display conformant to linear time.

Even though components of the Model are of an active, self-consistent nature, an additional flow generator needs to be present. Generating a flow in our context has to fulfil the task of resolving all open object dependencies, collecting the data and en passant performing co-ordinate transformations and at the core, linearising data with respect to time. As a result of such linear alignment, all Playlists are merged to form a complete script for the screenplay of the whole presentation. The flow generator as described is, if properly implemented, well suited for transmitting data collection of presentations as a sequential stream over the network.

## 4 Architecture and Implementation

### 4.1 Concepts

The technical core of the MobIT runtime environment is formed by an open multimedia architecture designed according to a 3-tiered principle as is shown in figure 4. A JAVA applet on the client side takes care of the presentation, processes the data object delivery and includes the flow generator by means of a sequentialising data interface and a time state machine. Media specific aspects are found encapsulated within element classes that control format processing and displaying independently. The element class requests implementations of methods according to the system's action types and can hence easily be extended. The client also provides the general option to receive and compose additional sub server classes and eventually run them in a separate thread.

The MobIT server is likewise written in JAVA and is primarily responsible for the session and transaction management with respect to client applets. Client and server communicate via a simple data exchange protocol that is spoken asynchronously through buffering cache layers allowing for latency hiding. To retrieve Media Data from a repository, the server relies on an abstract data interface, presently fed by the two data sources implemented. Media object structuring may be XML-coded for direct server processing to store data only in a flat file system. More elaborate data access is offered by an intelligent, multifunctional multimedia database system, which will be described in a forthcoming paper [9]. Both meta data and binary elements are readily handed over as JAVA object instances at the MobIT application server's demand.

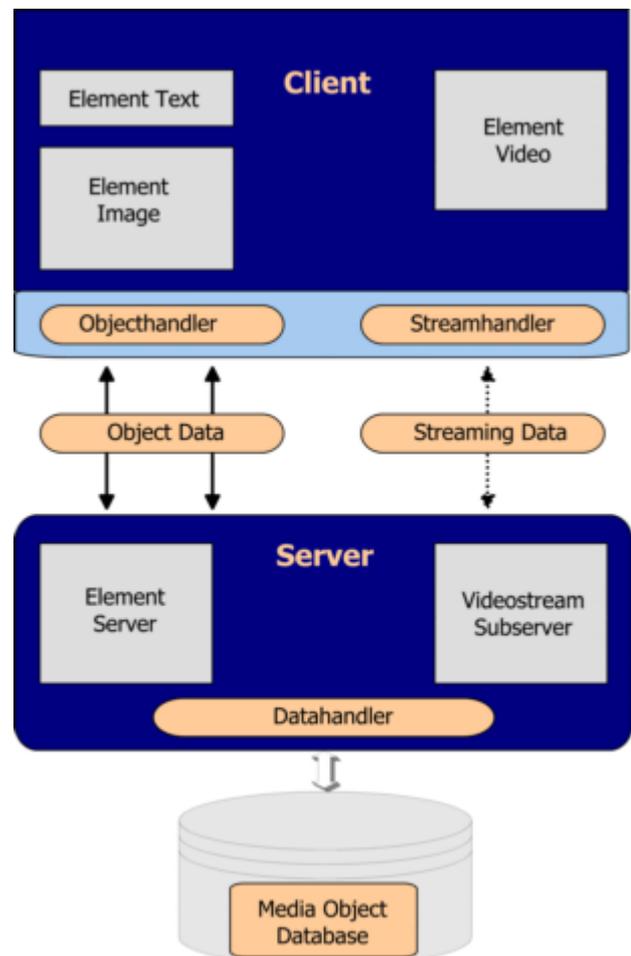

**Figure 4: System Architecture**

## 4.2 Sub serving

The ability to deal with pluggable sub servers may be seen as an important feature of this platform. Sub serving not only opens up the field for non-standard media and streams, but also allows for incorporation of new, complex functionalities such as online data processing without fattening the thin applet client. For an overall stream-oriented system, it appears quite natural to include served media for streaming. Mobil provides a flexible and simple interface for this purpose. Currently are used to incorporate the high performance optimised JAVA Wavelet video player sub servers of Cycon et al [2], a direct text sender that permits messaging to ongoing presentations and a LaTeX server that processes LaTeX formulas for display.

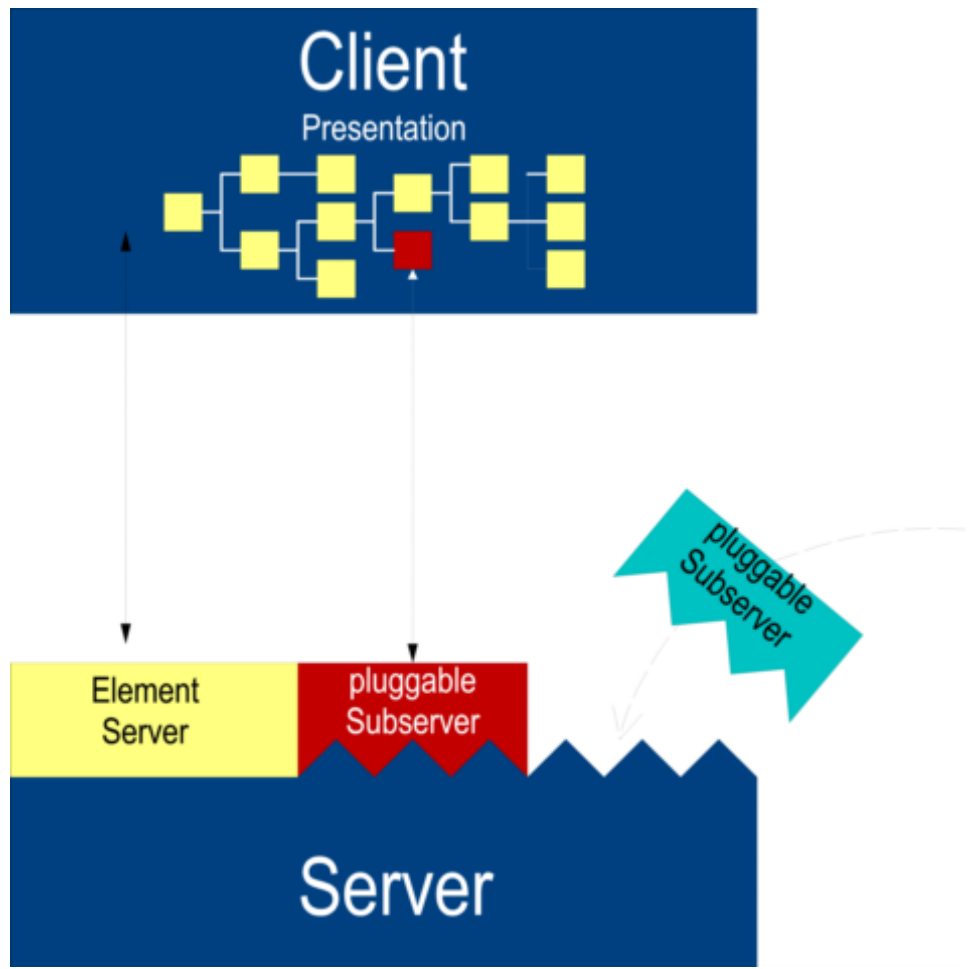

The interface to include any type of sub server has been purposely designed in a simple way: any sub server must implement the procedures `getPortCount` to allow for inquiry on requested number of ports, `setPorts` to permit port assignment, and `setData` to receive data handles and the initialisation. Additional information classes etc. are kept optional. The interface at the corresponding client site appears even simpler: `setServerInfo` and `getServerInfo` are the procedures needed here. Within this open framework, it should be easy to bring additional data servers to the system, for instance to include real-time visualization or live streams etc.

**Figure 5: Sub server Extensions**

## 4.3 Web Authoring

The system offers easy access for authors through a Web authoring tool. It is designed to guide authors through the different levels of complexity by means of several adapted views. Because it is well known, and to some extent obvious, that the WYSIWYG paradigm does not hold in the case of temporal, structural or event editing [14], we attempt to relate the multiple aspects of authoring to specific, intuitive appearances of the tool, thereby relying on the semantics of the structural relations of the system.

At the first stage of content authoring, our tool allows for data object upload and control. Guided by an object browser, the author may organise and retrieve objects in a directory structure of a virtual file system, assign names, media types and properties to the dobs and upload actual data to the data store (s. fig. 6). In general, media data manipulation is not meant to be part of the application authoring process, but for the sake of simplicity a simple text editor that also supports HTML-formatting is included in the system and permits the direct generation of written text.

Whereas the object browser in the MOB editing regime remains unchanged, dedicated support is given to the author in designing presentations. With the help of a structure view, a spatial view and an (relative) object timeline, authoring of MOB-based applications receives its basic tools. However, as was pointed out above, the specific semantic of media object structures is only fixed with the application layout. Thus the authoring requirements may change significantly between different fields of use and in general specific aspects cannot be foreseen. By including a toolbox of methods, our open system provides a programming interface to allow for easy, application-dependent extensions in the form of specific views.

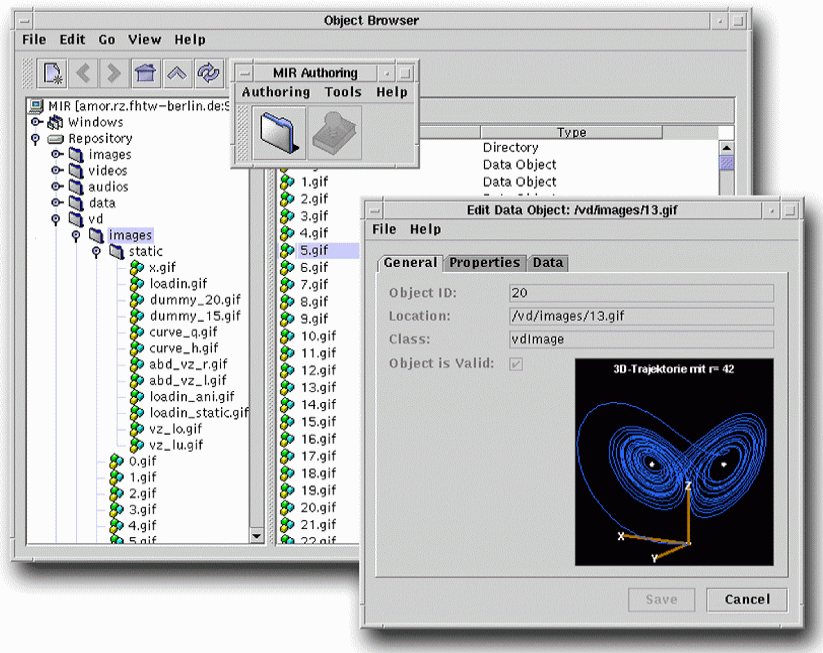

Figure 6: Authoring Media from the Web

## 5 Conclusions and Outlook

The lack of temporal aspects within today's attempts at Internet presentations and training has been recognized as a serious deficit. There are several interesting activities to overcome this. In the context of this debate, our paper presented the time-oriented teaching and presentation system MobIT that has been designed to handle complex, reusable media objects in time. We introduced a clear, straightforward concept that copes with temporal and spatial object hierarchies based on the CFM stream-oriented model. The prototypic implementation described here is still heavily under construction and it will be made publicly available at a more mature stage.

However, much work still has to be done in this ongoing project. Interactions have not yet been defined in CFM. As simple smil-type hyperlinks in our flow-oriented model could only support hopping between (possibly nested) parallel timelines, and as we do not intend to produce an interaction programming language, our current activities concentrate on modelling an interaction paradigm. Taking into account the CFM potential to operate on self-consistent media objects, we are aiming at a small 'alphabet' of operations, which enables authors to open up an unlimited number of navigational paths to the receptor with only a limited number of defined interactions.

Interactions will introduce an additional complexity to the treatment of network behaviour as they might contradict latency hiding technique. This is unavoidably true for user dialogue elements. For the loading of large binary elements, a careful time control will be needed. However the buffered pre-reading of Mobs may be viewed as filling an instruction cache of a processing unit. Interactions impose branches to the instruction flow and can be buffered in parallel so that immediate system response is generated.

## Acknowledgements

We would like to thank Hans Cycon and his group for the pleasant collaboration. They developed the highly optimised Wavelet Video algorithms and the codec. Mark Palkow carefully ported the codec to JAVA. Torsten Rack was responsible for the authoring tool. Special thanks go to Andreas Kárpáti who not only implemented the database system but also always lent an open ear or three in many fruitful discussions.

## Vitae:

**Björn Feustel**, born 1975, works at the computer centre of FHTW Berlin in the field of web-based multimedia. After completion of his traineeship in electronics, he studied computer sciences and meanwhile completed several projects focussing on web/database programming of virtual communities. He completed his diploma thesis on "Ein multimediales, zeitbasiertes Lehr- und Publikationssystem" which forms the basis of the current conference contribution.

**Thomas C Schmidt**, born 1964, is the head of the computer centre of FHTW Berlin. He studied mathematics and physics at the Freie Universität Berlin and the University of Maryland, USA. In 1993 he received his PhD in mathematical physics for work in many-particle theory quantum mechanics at the theory group of the Hahn-Meitner-Institut in Berlin. Since the late '80s he has been involved in many computing projects, especially focusing on simulation and parallel programming, distributed information systems and visualisation.